\documentclass[aps,showpacs,preprintnumbers,amsmath,amssymb]{revtex4}

\usepackage{balance}
\usepackage{hyperref} %
\usepackage{dcolumn}
\usepackage[dvips]{epsfig}

\usepackage{float}
\usepackage{graphicx}
\usepackage{epstopdf}
\usepackage{graphicx}
\usepackage{epstopdf}
\usepackage{color}
\usepackage{subfig}
\usepackage{amsmath,amssymb,amsfonts}

\begin{document}
\baselineskip=0.8 cm

\title{Thin accretion disk around a Kerr black hole immersed in swirling universes }
\author{Yuxin Ouyang$^{1}$, Xuan Zhou $^{3}$, Songbai Chen$^{1,2}$\footnote{Corresponding author: csb3752@hunnu.edu.cn} and Jiliang Jing$^{1,2}$\footnote{jljing@hunnu.edu.cn}}

\affiliation{$^1$Department of Physics, Institute of Interdisciplinary Studies, Hunan Research Center of the Basic Discipline for Quantum Effects and Quantum Technologies, Key Laboratory of Low Dimensional Quantum Structures
    and Quantum Control of Ministry of Education, Synergetic Innovation Center for Quantum Effects and Applications, Hunan
    Normal University,  Changsha, Hunan 410081, People's Republic of China
    \\
    $ ^2$Center for Gravitation and Cosmology, College of Physical Science and Technology, Yangzhou University, Yangzhou 225009, People's Republic of China
    \\
    $ ^3$School of Physics and Electronic Science, Hunan University of Science and Technology, Xiangtan 411021, China}

\begin{abstract}
\baselineskip=0.6 cm
\begin{center}
{\bf Abstract}
\end{center}

We have studied the properties of thin accretion disks around swirling-Kerr black holes, which own an extra swirling parameter describing the rotation of the immersed universe. Our results show that the swirling parameter leaves distinct imprints on the energy flux, temperature distribution and emission spectra of the disk and  gives rise to some new effects that differ from those induced by  the black hole's spin. With the increasing of the swirling parameter, both the energy flux and  radiated temperature in the disk increase in the inner region where circular orbital radii are smaller  and decrease in the  outer region where circular orbital radii are larger. In contrast, these quantities consistently increase with the black hole's spin. Although  the swirling parameter and the black hole's spin parameter lead to  higher cut-off frequencies, the background swirling  reduces the observed luminosity of the disk at lower frequencies and enhances it only at higher frequencies, which is quite distinct from that of the black hole's spin. Furthermore, the conversion efficiency increases with the black hole's spin parameter, but decreases with the swirling parameters. Additionally, the effects of the swirling parameter are found to be suppressed by the black hole's spin parameter. These results could help us further understand the properties of thin accretion disks and the swirling of the universe background.
\end{abstract}
\maketitle
\newpage

\section{Introduction}

The accretion disk is a structure formed by the diffuse material gradually spiraling inward due to the gravitational field of a central compact body, which is an important research topic in the astrophysics.  Actually, accretion around a real celestial body is a highly complex dynamic process, which is comprehensively described by high-precision numerical simulations requiring exascale computational resources \cite{CCP,KBH,IGR,HAC,LCA}. Many theoretical models have been developed to study accretion disks, among which the steady-state thin accretion disk model is one of the simplest \cite{SOA}. In this model, the disk's thickness is negligible, allowing the heat generated by internal stress and dynamic friction to be radiated away through its surface. This efficient cooling mechanism ensures the maintenance of the disk's hydrodynamic equilibrium and a time-independent constant mass accretion rate.  The physical properties of steady-state thin accretion disks in various background spacetimes have been investigated extensively in \cite{SMG,ADP,IWG,AWS,FBH,WNS,AQS,BCO,BFS,BHC,WBH,MGM,DBH,SOA,CCP,KBH,IGR,HAC,LCA,OAD,OTH}. The impacts of magnetic fields and a quadrupole moment on the spectral characteristics and properties of thin accretion disks around a distorted black hole have been explored in \cite{DBH}. Quantum corrections have been examined for their impacts on the observed radiation fluxes from thin accretion disks \cite{CBH}. The properties of thin accretion disks have been investigated in  the backgrounds of short-hairy black holes \cite{HBH} and rotating hairy black holes \cite{AOA}, as well as quark stars \cite{WNS1}, boson stars \cite{BCO1}, and fermion stars \cite{BFS1}. The unique signatures in  energy flux and  emission spectrum of the disk can provide  valuable insights into the central celestial bodies but also offer a profound means to test alternative theories of gravity.

Since all compact objects in the universe are rotating, it is natural that the rotation is a
universal phenomenon which may also be applied to the global universe. Recently, a kind of non-trivial solutions immersed in a swirling universe are generated by applying the Ehlers transformation to the Ernst potenials for the magnetic form of the Lewis-Weyl-Papapetrou (LWP) metric. For stationary and axisymmetric spacetimes, the Einstein-Maxwell equations are equivalent
to the complex Ernst equations \cite{GFP,GFP1}
\begin{eqnarray}
(Re\;\mathcal{E}+|\Phi|^2)\nabla^2\mathcal{E} =\nabla\mathcal{E}\cdot(\nabla\mathcal{E}+2\Phi^{*}\nabla\Phi),\quad\quad(Re\;\mathcal{E}+|\Phi|^2)\nabla^2\Phi =\nabla \Phi\cdot(\nabla\mathcal{E}+2\Phi^{*}\nabla\Phi),\label{Ernstform}
\end{eqnarray}
where $\mathcal{E}$ and $\Phi$ are complex functions representing the Ernst
potentials. The operator $\nabla$ is the flat vectorial operator in Euclidean space with cylindrical coordinates $(\rho,\;\varphi\;z)$.
The magnetic LWP metric has a  form
\begin{eqnarray}
ds^2=f^{-1}[-\rho^2dt^2+e^{2\gamma}(d\rho^2+dz^2)]+f(d\varphi-\omega dt)^2,\label{LWPmetric}
\end{eqnarray}
where $f$, $\omega$ and $\gamma$ are functions of $\rho$ and $z$ only. Obviously, the metric given by Eq. (\ref{LWPmetric}) is stationary and axisymmetric, implying that its Ernst equations take the same form as presented in Eq. (\ref{Ernstform}).
The corresponding Ernst potentials are
\begin{eqnarray}
\mathcal{E}=-f-|\Phi|^2+ih,\quad\quad\quad \Phi=\tilde{A}_t-iA_{\varphi},
\end{eqnarray}
where  the twisted potentials $\tilde{A}_t$ and $h$ respectively satisfy
\begin{eqnarray}
\hat{e}_{\varphi}\times\nabla\tilde{A}_t=-\frac{f}{\rho}\bigg(\nabla A_t+\omega\nabla A_{\varphi}\bigg),\quad\quad\quad \hat{e}_{\varphi}\times\nabla h=-\frac{f^2}{\rho}\nabla\omega-2\hat{e}_{\varphi}\times Im(\Phi^{*}\nabla \Phi).
\end{eqnarray}
Here $A_t$ and $A_{\varphi}$ are electromagnetic potential components in the spacetime.
It is well known that the Ernst equations (\ref{Ernstform}) are invariant under the Ehlers and
the Harrison transformations.  The Ehlers transformation of the Ernst potenials is
\begin{eqnarray}
\mathcal{E}\rightarrow \mathcal{E}'=\frac{\mathcal{E}}{1+ij\mathcal{E}},\quad\quad\quad \Phi\rightarrow \Phi'=\frac{\Phi}{1+ij\mathcal{E}}.\label{pottransformation1}
\end{eqnarray}
For the chosen seed metric, applying the Ehlers transformation (\ref{pottransformation1}) allows one to generate the corresponding swirling spacetime. This generated swirling spacetime solution is non-trivial, because the Ehlers map belongs to $SU(2, 1)$ rather than Abelian group \cite{AbeMiso}.
When the Minkowski metric is selected as the seed, the aforementioned operations yield a solution for a swirling universe, which describes the geometry of a rotating background universe.
 Such a swirling universe can be interpreted as gravitational vortices created by a pair of counter-rotating sources situated at infinite distance \cite{BHS}. In cylindrical coordinates, the form of the swirling universe can be simplified as
\begin{eqnarray}
ds^2=(1+j^2\rho^4)(-dt^2+d\rho^2+dz^2)+\frac{\rho^2}{1+j^2\rho^4}(d\varphi+4jzdt)^2,
\end{eqnarray}
 and the symmetries of the metric are defined by the four independent Killing vectors $(\partial_t,\;\partial_{\varphi},\;z\partial_t+t\partial_z-2j(t^2+z^2)\partial_{\varphi},\; \partial_z-4jt\partial_{\varphi})$ and the non-trivial Killing-Yano 2-form $-4j\rho zdt\wedge d\rho+j\rho^2(1+j^2\rho^4)dt\wedge dz+\rho d\rho\wedge d\varphi$.
 The swirling universe spacetime  belongs to the Petrov type $D$ class \cite{BHS}, with its corresponding Newman-Penrose spin coefficients all vanishing. Furhtermore, the swirling universe spacetime  is everywhere regular since the Kretschmann scalar $R_{\mu\nu\rho\sigma}R^{\mu\nu\rho\sigma}$ is finite. Especially, there are no closed timelike curves, and related causality issues are absent, since curves with constant $t$, $\rho$ and $z$ satisfy $ds^2=\frac{\rho^2}{1+j^2\rho^4}d\varphi^2>0$.
Starting from a Kerr seed spacetime and exploiting the Ernst formalism \cite{GFP,GFP1,AMU,AMU1,OTC} in combination with the Ehlers transformation, Astorino \textit{et al} \cite{ASU1} obtained an exact solution for a rotating black hole immersed in a swirling universe. It is a vacuum solution of Einstein field equations and has an extra  swirling parameter that describes the rotation of background universe. The spin-spin interaction between the black hole and the background frame dragging results in a peculiar $Z_{2}$ symmetry with the northern and southern hemispheres rotating in opposite directions, and the non-removable conical singularities along the axis of symmetry. The black hole horizons are located at $r_{\pm}=M\pm \sqrt{M^2-a^2}$, which is equivalent to the Kerr case, but the swirling background deforms the horizon geometry \cite{ASU1}. The geodesic equations in the swirling universe are found to no longer be decoupled, leading to chaotic motion of particles in this spacetime \cite{SOS,ISU1}. Moreover, the swirling parameter is found to cause the shadow's contour to take on a tilted oblate shape\cite{ISU}. Additionally, the properties of the geometrically thick equilibrium tori have been investigated in the backgrounds of  a Schwarzschild black hole in swirling universe \cite{disk1} and a Kerr black hole in swirling  universe \cite{ASU}, respectively. The studies on the swirling black holes have been generalized to the electromagnetic case \cite{SMU,TMC} and to the equilibrium configuration  of two rotation black holes \cite{SSI}. Recently, adopting a composition of magnetic Ehlers and Harrison transformations, several new swirling spacetimes are obtained and their  spacetime features are investigated \cite{TIB}. Furthermore, a new rotating vacuum solution of Einstein equations has been obtained \cite{LCS}, which has a swirling asymptotic behaviour, while possessing a well-defined static limit. These studies shed new light on understanding black holes in a swirling universe.
The main motivation of this paper is to study the properties of a thin accretion disk around a swirling-Kerr black hole and to probe effects arising from  the background swirling on the disk's energy flux, temperature distribution and emission spectra of the disk, and  to see what new effects differed from those induced by the black hole's spin.

The plan of this paper is as follows: In the Sec.II, we briefly review the swirling-Kerr black hole solution \cite{ASU1}  and  the timelike geodesics in this spacetime background. In Sec.III, we present properties of thin accretion disks in the swirling background and probe effects of both the swirling parameter and the black hole spin parameter. Finally, we present a summary.

\section{Geodesic Motion Of Test Particles around a Kerr black hole immersed in a swirling universe}

Let us now review briefly the Kerr black hole solution immersed in a swirling universe background. It belongs to a family of stationary, axially symmetric and non-asymptotically flat black hole solutions satisfying the vacuum Einstein equations. The swirling Kerr black hole solution can be obtained by utilizing a solution-generating technique and Ehlers transformation on the Kerr solution as a seed \cite{ASU1}. This rotating black hole possesses an extra swirling parameter that describes the swirling of universe background. In the Boyer-Lindquist coordinates, the metric form of this swirling black hole can be written as \cite{ASU1}
\begin{equation}\label{metric}
    \begin{aligned}
        ds^{2}=F(d\phi-\omega dt)^{2}+F^{-1}[-\rho^{2} dt^{2}+\Sigma\sin^{2}\theta(\frac{dr^{2}}{\Delta}+d\theta^{2})],
    \end{aligned}
\end{equation}
with
\begin{equation}
    \begin{aligned}
        F^{-1}=\chi_{(0)}+j\chi_{(1)}+j^{2}\chi_{(2)}, \quad \quad \omega=\omega_{(0)}+j\omega_{(1)}+j^{2}\omega_{(2)}.
    \end{aligned}
\end{equation}
Here functions $F$ and $\omega$ are expanded in a finite power series of $j$, and the corresponding  expansion coefficients $\chi_{(i)}$ and $\omega_{(i)}$ are
\begin{equation}
\chi_{(0)}=\frac{R^2}{\Sigma \sin^2\theta},\quad\quad\quad \chi_{(1)}=\frac{4aM\Xi\cos{\theta}}{\Sigma \sin^{2}\theta},\quad\quad\quad
\chi_{(2)}=\frac{4a^{2}M^{2}\Xi^{2}\cos^2{\theta}+\Sigma^{2}\sin^{4}{\theta}}{R^2\Sigma \sin^2{\theta}},
\end{equation}
and
\begin{eqnarray}
        \omega_{(0)}&=&\frac{2aMr}{-\Sigma},\quad\quad\quad\omega_{(1)}=\frac{4\cos{\theta}[-a\Omega(r-M)+Ma^4-r^4(r-2M)-\Delta a^2 r]}{-\Sigma},\nonumber\\
        \omega_{(2)}&=&-\frac{2M}{\Sigma}\bigg\{3ar^5-a^5(r+2M)+2a^3r^2(r+3M)-r^3(\cos^2{\theta}-6)\Omega\nonumber\\
        &+&a^2[\cos^2{\theta}(3r-2M)-6(r-M)]\Omega\bigg\},
\end{eqnarray}
where
\begin{eqnarray}
\Delta&=&r^2-2Mr+a^2,\quad \quad \quad \rho^2=\Delta\sin^2{\theta},\quad \quad \quad \Sigma=(r^2+a^2)^2-\Delta a^2\sin^2{\theta},  \nonumber \\
\Omega&=&\Delta a \cos^2{\theta},\quad \quad \quad  R^2=r^2+a^2\cos^2{\theta},\quad \quad \quad \Xi=r^2(\cos^2{\theta}-3)-a^2(1+\cos^2{\theta}).
\end{eqnarray}
Here parameters $M$ and $a$ respectively correspond to the mass parameter and the specific angular momentum of the black hole. This swirling effect endows the black hole with novel and distinct properties that differ from those of the pure Kerr black hole. For example, a conical singularity along the symmetry axis \cite{SSI}, caused by the spin-spin interaction between the black hole and the background dragging, deforms the horizon geometry and enhances the symmetry breaking of the spacetime properties. The north and south hemispheres of the swirling Kerr black hole rotate in opposite directions. It reduces to the pure Kerr black hole as the swirling parameter $j = 0$ and to the Schwarzschild black hole in swirling universes as the black hole spin $a =0$. If the parameters $M=0$ and $a=0$, the black hole disappears and there is only a swirling background left. It must be pointed out that in the swirling black hole spacetimes one cannot define an everywhere timelike Killing vector and consequently a privileged
zero angular momentum observer (ZAMO) due to the presence of ergoregions that extended all the way to infinity \cite{TIB,LCS}. For the spacetime (\ref{metric}), the  metric component $g_{tt}$
be written in the following form \cite{KerrNASA}
\begin{eqnarray}
g_{tt}=g_{(1)}\sin^2\theta+g_{(2)}\Delta,
\end{eqnarray}
with
\begin{eqnarray}
g_{(1)}&=&\frac{\omega^2\Sigma}{\mathcal{B}},\quad\quad \quad g_{(2)}=-\frac{\mathcal{B}}{\Sigma},
\quad\quad \quad\mathcal{B}=R^2+4ajM\Xi\cos{\theta}+4a^{2}j^2M^{2}\Xi^{2}\cos^2{\theta}+j^2\Sigma^{2}\sin^{4}{\theta}.
\end{eqnarray}
Obviously, both $g_{(1)}$ and $g_{(2)}$ are differentiable functions. Furthermore, it can be proved that on the symmetry axis $(\sin^2\theta=0)$ it holds $g_{(2)}<0$. Therefore, the part of symmetry axis outside the outer horizon does not belong to the ergoregion since $g_{tt}<0$, which means that the Killing vector $\partial_t$ is timelike  and one can define a ZAMO in this special region. Base on this point, we restrict the observer to the axis of symmetry in our analysis.

For the thin accretion disk model, one can assume that the disk is on the equatorial plane and that the matter moves on nearly geodesic circular orbits. In the swirling Kerr black hole spacetime, the timelike geodesic equations of a particle moving on the equatorial plane $\theta=\pi/2$ can be
expressed as
\begin{eqnarray}
u^t&=&\frac{F(E+L\omega)}{\rho^2}, \quad \quad \quad
u^{\phi}=\frac{L}{F}-\frac{F\omega(E+L\omega)}{\rho^2},\nonumber\\
g_{rr}\bigg(\frac{dr}{d\tau}\bigg)^{2}&=&\frac{g_{\phi\phi}E^{2}+2g_{t\phi}EL+g_{tt}L^{2}}{g^2_{t\phi}-g_{tt}g_{\phi\phi}}-1\equiv V_{eff}.
\end{eqnarray}
Here $E$ and $L$ correspond to the energy and the angular momentum of the timelike particle, respectively. The function $V_{eff}$ is the corresponding effective potential.
\begin{figure}[ht]
    \centering
\includegraphics[width=0.65\linewidth]{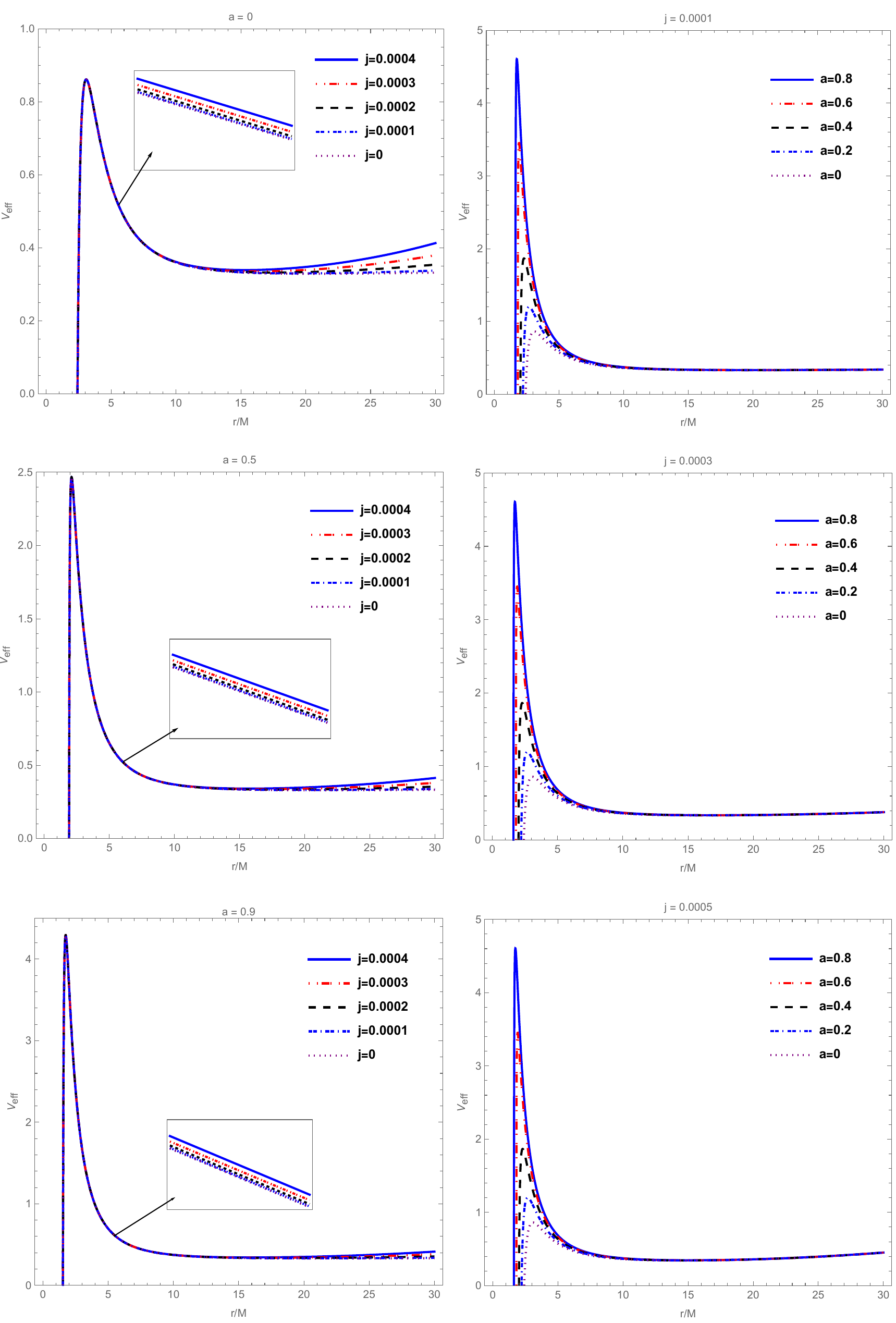}
\captionsetup{justification=raggedright, singlelinecheck=false}
    \caption{Changes of the effective potential $V_{eff}$ with the swirling parameter $j$ and the spin parameter $a$. Here, we set $M=1$ and  $E=0.8$ and $L=4$.}
  \label{Veff}
\end{figure}
Fig. (\ref{Veff}) illustrates that the effective potential increases with the swirling parameter $j$ as well as the black hole spin parameter $a$. The similar dependence of the effective potential on $j$ and $a$ is understandable since both parameters describe the rotation of spacetimes.

\begin{figure}[htbp]
    \centering
		\includegraphics[width=0.86\linewidth]{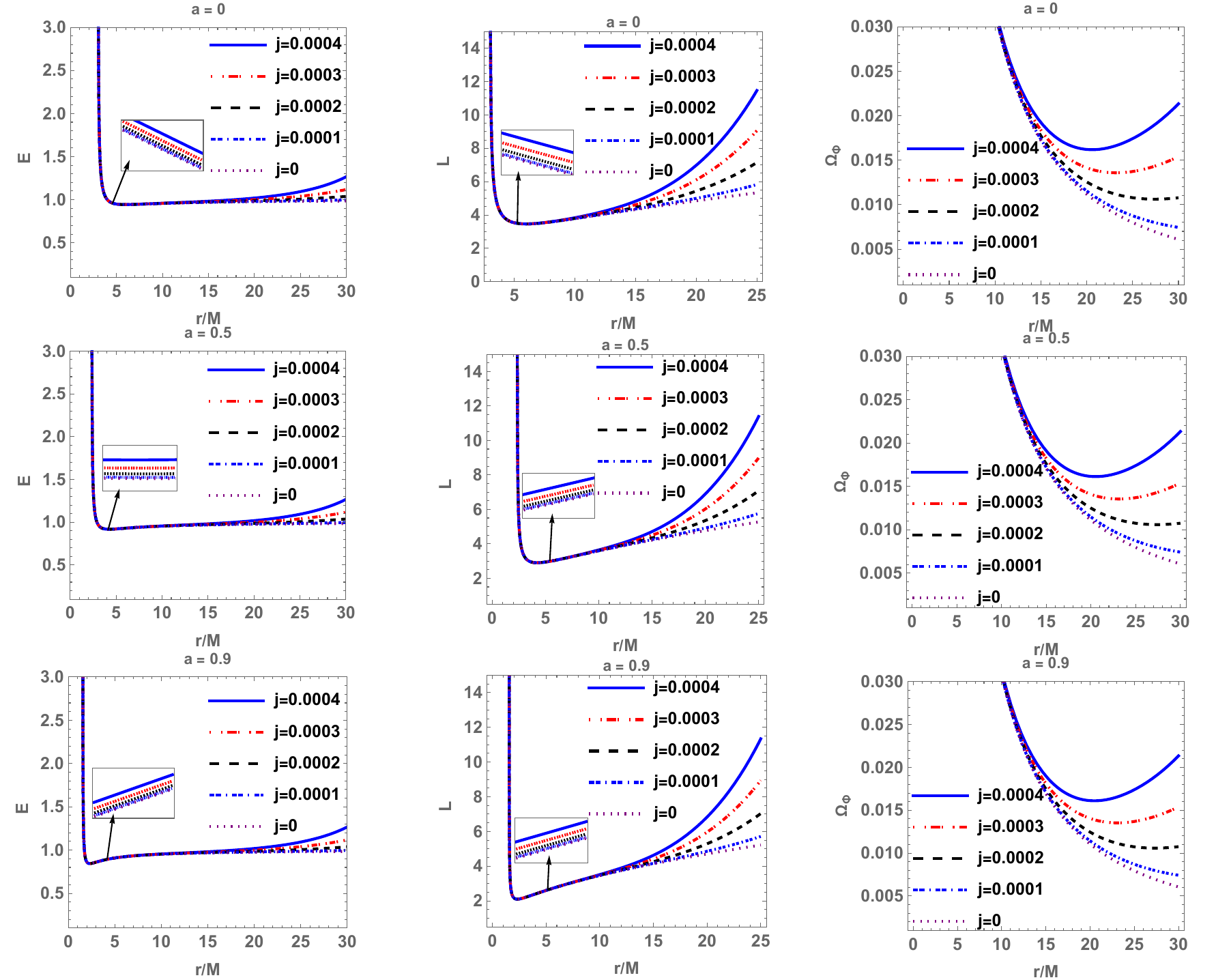}
		\includegraphics[width=0.86\linewidth]{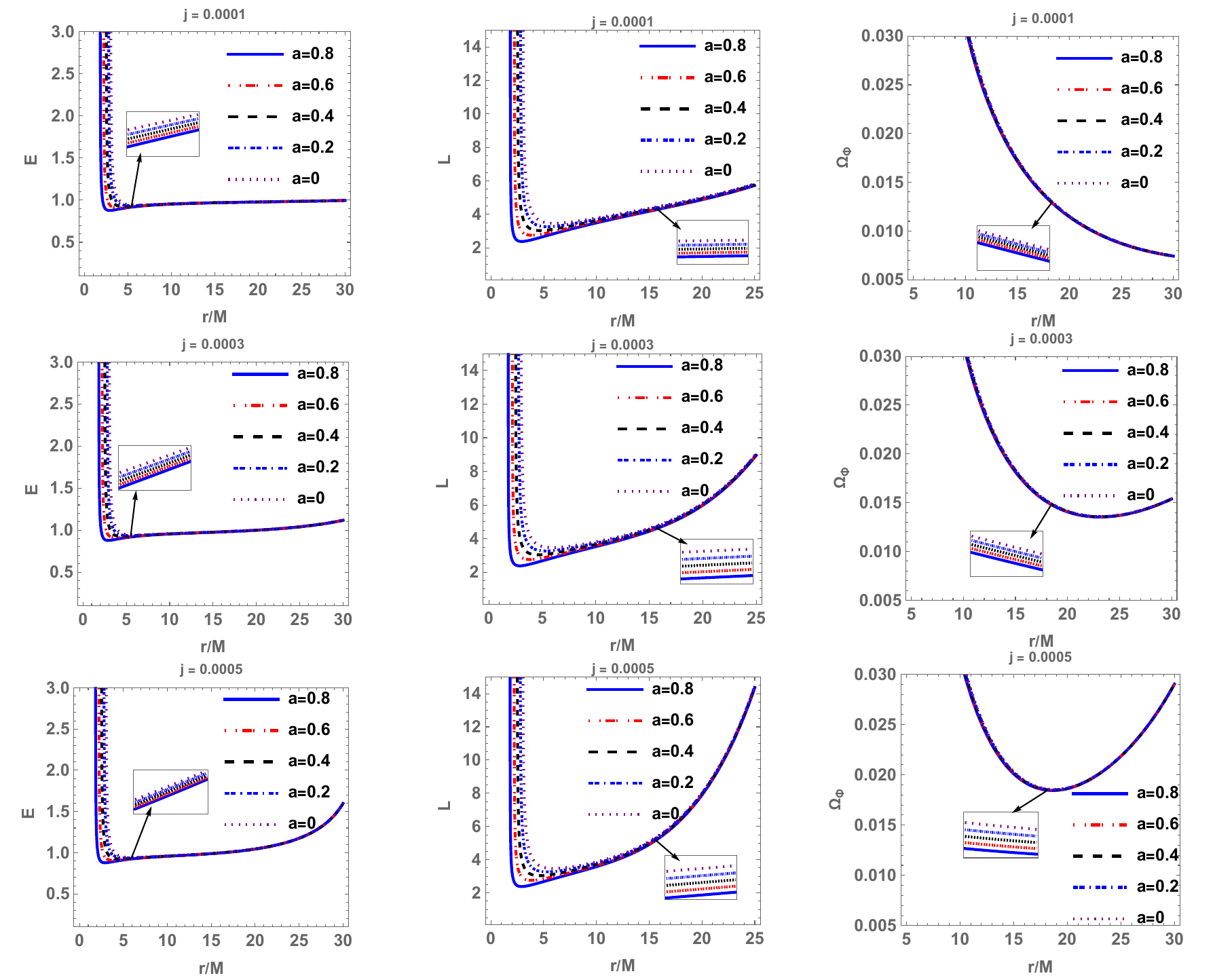}
\captionsetup{justification=raggedright, singlelinecheck=false}
 	\caption{Changes of $E$, $L$ and $\Omega_{\phi}$ of particles moving along circular orbits with the swirling parameter $j$ and spin parameter $a$. Here we set $M=1$. }
        \label{ELW}
\end{figure}

\begin{figure}[htbp]
    \centering \includegraphics[width=0.43\linewidth]{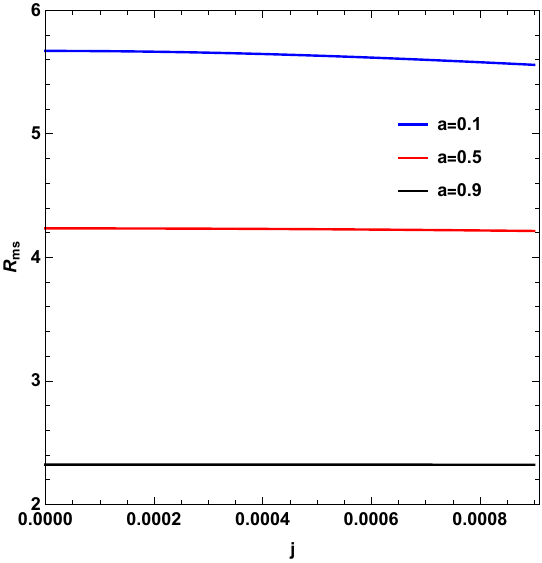}\;\;\;\;\;\includegraphics[width=0.44\linewidth]{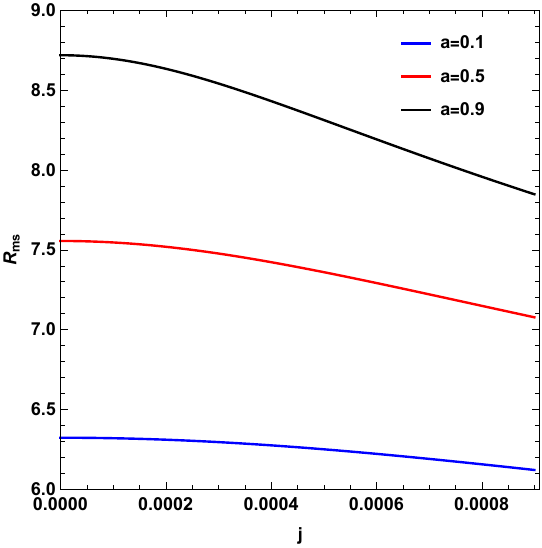}
\captionsetup{justification=raggedright, singlelinecheck=false}
 	\caption{Changes of the marginally stable orbit radius $r_{ms}$ with the parameter $j$ for the thin disk around the swirling Kerr black hole. The left panels correspond to the prograde case, and the right panels correspond to the retrograde case.}
        \label{rms}
\end{figure}

On the equatorial plane, the particles moving along circular orbits satisfy the conditions $V_{eff}=0$ and $V_{eff,r}=0$. Making use of these conditions, we can get the specific energy $E$, the specific angular momentum $L$, and the angular velocity $\Omega_{\phi}$ of the particle moving in circular orbit on the equatorial plane in the swirling Kerr black hole spacetime
\begin{eqnarray}
&&E=-\frac{g_{tt}+g_{t\phi}\Omega_{\phi}}{\sqrt{-g_{tt}-2g_{t\phi}\Omega_{\phi}
-g_{\varphi\varphi}\Omega^2_{\phi}}},\nonumber\\
&&L=\frac{g_{t\phi}+g_{\phi\phi}\Omega_{\phi}}{\sqrt{-g_{tt}
-2g_{t\phi}\Omega_{\phi}-g_{\phi\phi}\Omega^2_{\phi}}},\nonumber\\
&&\Omega_{\phi}=\frac{d\phi}{dt}=\frac{-g_{t\phi,r}\pm \sqrt{g^2_{t\phi,r}-g_{tt,r}g_{\phi\phi,r}}}{g_{\phi\phi,r}}.\label{jsd}
\end{eqnarray}
The signs $+$ and $-$ in $\Omega_{\phi}$ respectively correspond to prograde and retrograde orbits. In Fig. (\ref{ELW}), we find that for a fixed circular orbital radius,  the specific energy $E$,  specific angular momentum $L$, and  angular velocity $\Omega_{\phi}$ of the particle increase with the swirling parameter $j$, but decrease with the black hole spin parameter $a$. Especially, with the increasing of the orbital radius,  the angular velocity $\Omega_{\phi}$ in the non-zero $j$ case first decreases and then increases, which is different from that in the pure Kerr black hole case where it decreases monotonically.
Moreover,  the influence of the swirling parameter is more distinct in regions with larger $r$, while the effect of the spin parameter is stronger in regions with smaller  $r$. This means that the swirling of the universe background could give rise to some new effects differed from those arising from the black hole spin.
For particles moving on the equatorial plane, the marginally stable orbit radius is determined by the condition $V_{eff,rr}=0$. Unfortunately, in the swirling Kerr black hole spacetime (\ref{metric}), one cannot obtain an analytical form for the marginally stable orbit radius and must resort to numerical methods. In Fig. (\ref{rms}), we plot the variation of the marginally stable orbit radius $r_{ms}$ with the swirling parameter $j$ and the black hole spin parameter $a$, which shows that  $r_{ms}$ decreases with swirling parameter $j$ both in the  prograde and  retrograde cases for different values of $a$.
For a fixed $j$, it is easy to find that $r_{ms}$  decreases with the black hole spin parameter $a$ in the prograde case, but increases with $a$ in the retrograde case. This means that the effects of the swirling parameter on the marginally stable orbit radius $r_{ms}$  are different from those of the black hole spin parameter $a$.

\section{Properties of thin accretion disks around a Kerr black hole in swirling universes}

In this section we will study the accretion process in the thin disk around the swirling Kerr black hole by using the steady-state thin accretion disk model \cite{IBH} and probe effects of the swirling parameter $j$ on the energy flux,  conversion efficiency,  radiation temperature and spectra  in the disk.
In steady-state accretion disk models, the maximum half-thickness of the thin disk $H$ is assumed to be much smaller than its characteristic radius $r$ ($H \ll r$),  and the mass accretion rate $\dot{M_{0}}$ is assumed to remain a constant. Moreover, the accreting
matter in the disk is also assumed to  be described by an anisotropic fluid with the energy-momentum tensor \cite{IBH}
\begin{equation}\label{EM}
    \begin{aligned}
        T^{\mu\nu}=\rho_{0}u^{\mu}u^{\nu}+2u^{(\mu}q^{\nu)}+t^{\mu\nu},
    \end{aligned}
\end{equation}
where the density $\rho_{0}$,  energy flow vector $q^{\mu}$ and  stress tensor $t^{\mu\nu}$ are measured in the averaged rest-frame of the orbiting particle with four-velocity $u^{\mu}$. In the averaged rest-frame, the equations $u_{\mu}q^{\mu}=0$ and $u_{\mu}t^{\mu\nu}=0$ are satisfied.
The plasma of this thin disk satisfies the four dimensional conservation laws of rest mass,  energy and  angular momentum
\begin{equation}
    \begin{aligned}
      \nabla_{\mu}(\rho_{0}u^{\mu})=0,\quad\quad\quad
       \nabla_{\mu}E^{\mu}=0,\quad\quad\quad
       \nabla_{\mu}J^{\mu}=0,
    \end{aligned}
\end{equation}
where $E^{\mu}$ and $J^{\mu}$ represent the four-vectors of  energy and angular momentum flux, respectively. In the background of a rotating and swirling black hole, one can find
that the time-averaged radial structure equations of the thin disk
can be expressed as
\begin{eqnarray}
       &&\dot{M_{0}}=-2\pi r\Sigma(r)u^{r}=const, \\
        && \dfrac{d} {dr}(\dot{M_{0}}E-2\pi r\Omega_{\phi} W_{\phi }^{\phantom{\phi} r})=4\pi\sqrt{-g}KE,\label{EC} \\
        && \dfrac{d} {dr}(
    \dot{M_{0}}L-2\pi rW_{\phi }^{\phantom{\phi} r})=4\pi\sqrt{-g}KL,\label{LC}
\end{eqnarray}
where the averaged rest mass density $\Sigma(r)$ and the averaged torque $W_{\phi }^{\phantom{\phi} r}$ are
\begin{eqnarray}
    \Sigma(r)=\int_{-H}^{H}\langle\rho_{0}\rangle dz,\quad\quad\quad   W_{\phi}^{\phantom{\phi} r}=\int_{-H}^{H}\langle t_{\phi}^{\phantom{\phi} r}\rangle dz.
\end{eqnarray}
Here $\langle t_{\phi}^{\phantom{\phi} r}\rangle$ refers to the average
value of the $\phi$-$r$ component of the stress tensor over a characteristic
time scale $\Delta t$ and the azimuthal angle  $\Delta\phi=2\pi$. By eliminating $W_{\phi}^{\phantom{\phi} r}$ from Eqs. (\ref{EC}) and (\ref{LC}) and applying the energy-angular momentum relation  for  circular geodesic orbits, $E_{,r}=\Omega_{\phi} L_{,r}$, one can find that the energy flux $K(r)$ can be expressed as
\begin{equation}
    \begin{aligned}
        K(r)=-\frac{\dot{M_{0}}\Omega_{\phi,r}}{4\pi\sqrt{-g}(E-L\Omega_{\phi})^{2}}\int_{r_{ms}}^{r}(E-L\Omega_{\phi})L_{,r} dr.
    \end{aligned}
\end{equation}
\begin{figure}[ht]
    \centering \quad\quad\includegraphics[width=0.32\linewidth]{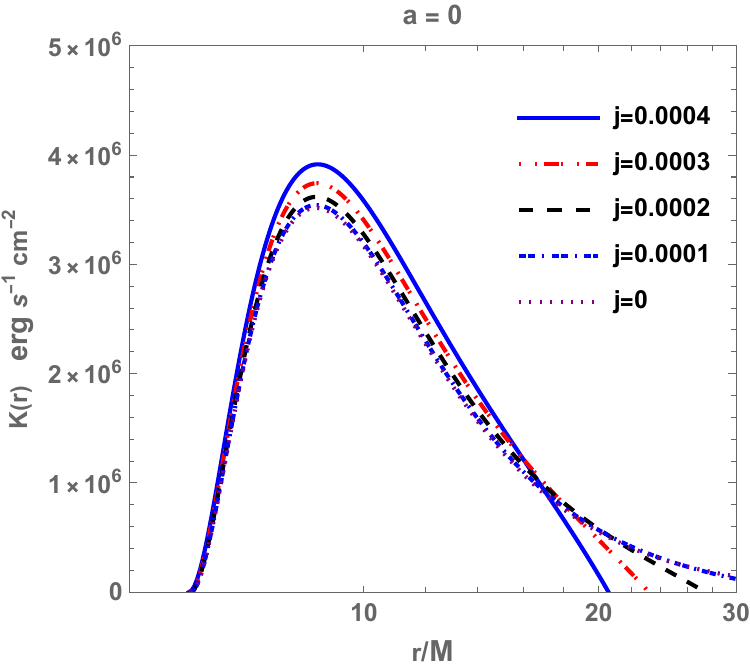}\includegraphics[width=0.35\linewidth]{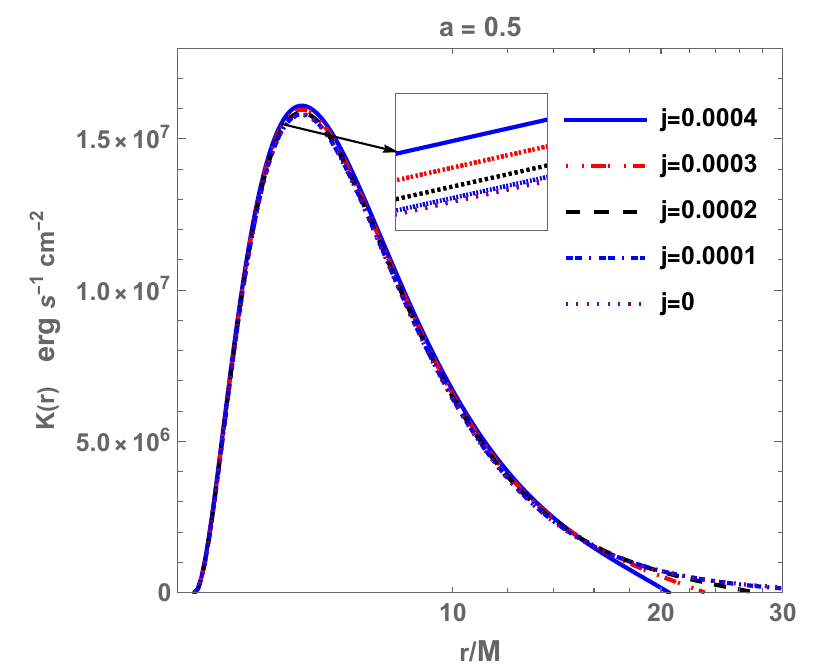}\includegraphics[width=0.35\linewidth]{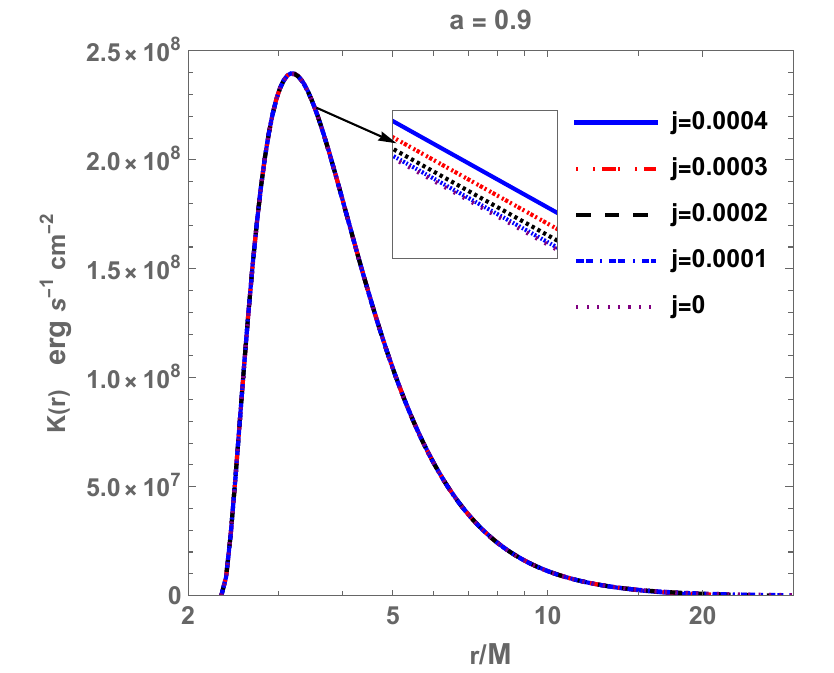}\\
     \includegraphics[width=0.32\linewidth]{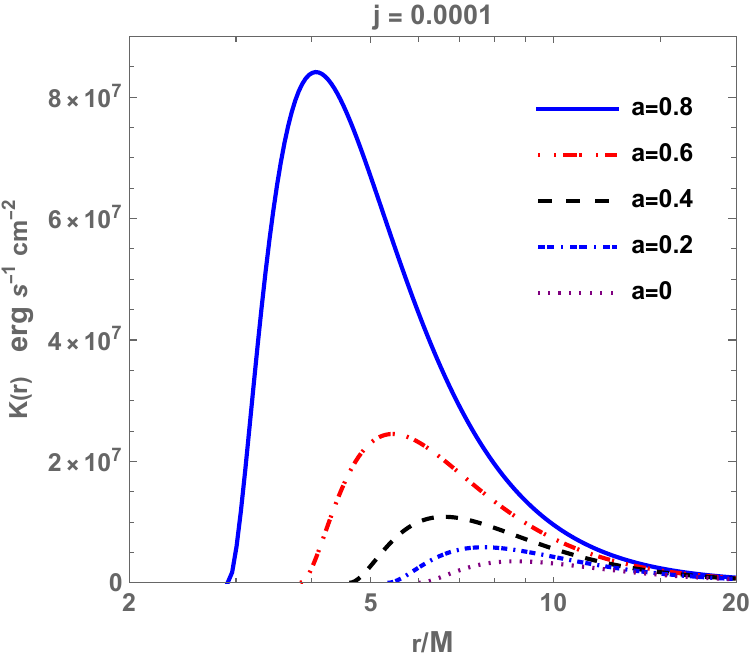}\includegraphics[width=0.35\linewidth]{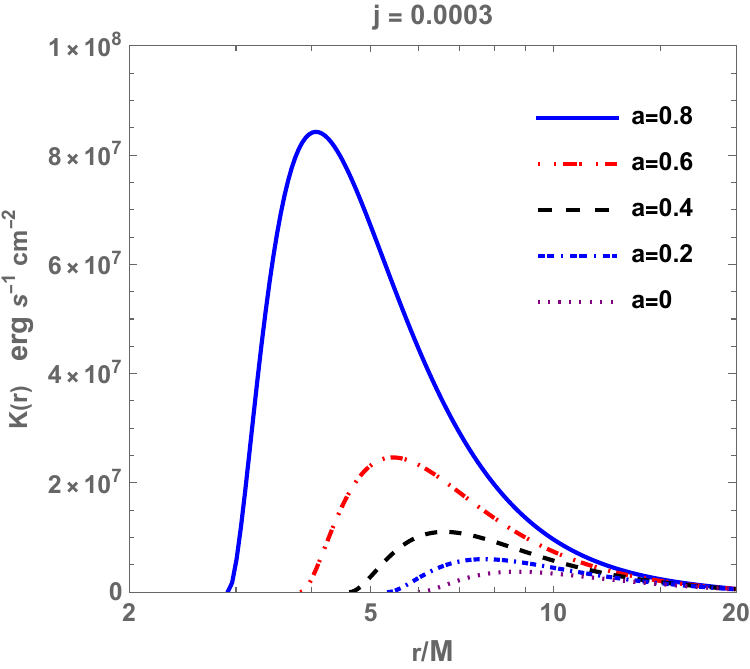}\includegraphics[width=0.35\linewidth]{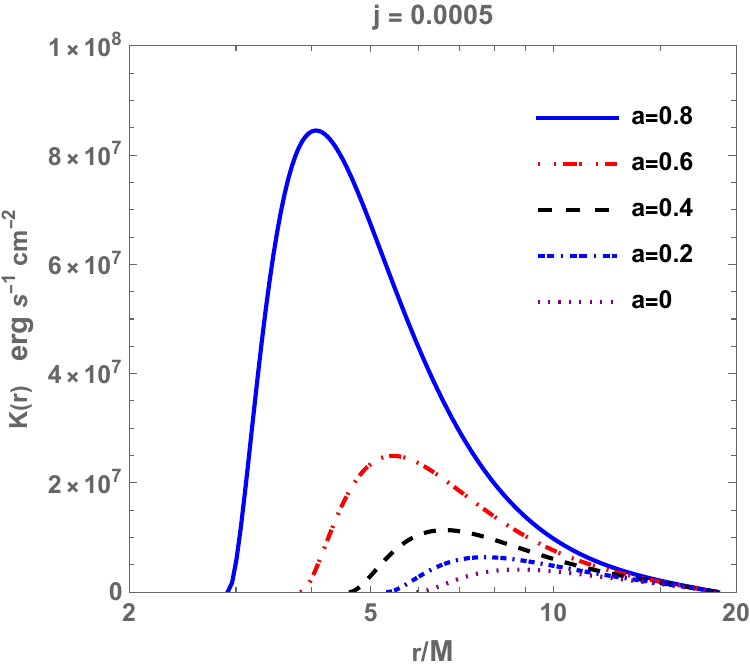}
\captionsetup{justification=raggedright, singlelinecheck=false}
 	\caption{Variation of the energy flux $K(r)$ with the swirling parameter $j$  and the black hole spin parameter $a$  in the thin disk around a rotating swirling black hole. Here, we set $M=1$.}\label{K}
\end{figure}
Let us now consider the mass accretion driven by black holes with a total mass of $M=10^6 M_{\odot}$ and a mass accretion rate $\dot{M}_{0}=10^{-12}M_{\odot} yr^{-1}$, these parameters are also adopted in the analysis of the thin accretion disk model around non-commutative Kerr black holes, as presented in \cite{KBH1}. In Fig. (\ref{K}), we present the total energy flux radiated by a thin disk around the swirling Kerr black hole for different values of the swirling parameter and spin parameter.
As shown in Fig. (\ref{K}), the energy flux $K$ increases with the black hole spin $a$ for different $j$, which is similar to that in the Kerr black hole case. With the increasing of the swirling parameter $j$, the energy flux $K$ increases in the inner region with the smaller circular orbital radius and decreases in the outer region with the larger circular orbital radius. Moreover, in the left panel of Fig. (\ref{K}), we also note that in the case $j \neq 0 $, the energy flux $K(r)$ becomes zero at the orbit where the radius meets $\Omega_{\phi,r}=0$, which does not appear in the Kerr case. When $\Omega_{\phi,r}>0$ and the radiated energy flux $K<0$, which implies that in this case the particle moving along the circular orbit is excreted from the disk to the spatial infinity rather than being accreted into the black hole. Thus,  the accretion disk model is not valid in this scenario,  so we focus on only the case where $\Omega_{\phi,r}<0$.
\begin{figure}[ht]
    \centering \includegraphics[width=0.35\linewidth]{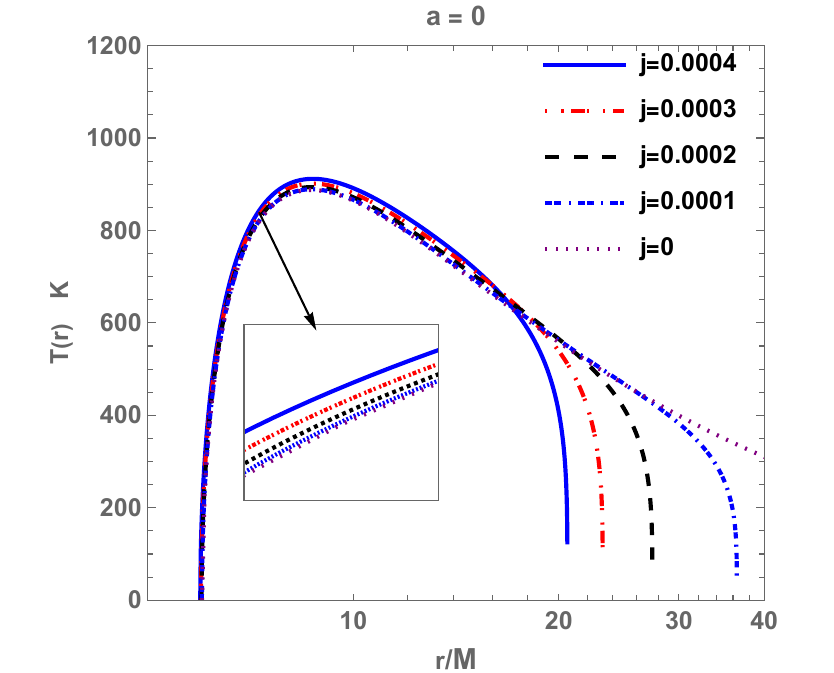}\includegraphics[width=0.34\linewidth]{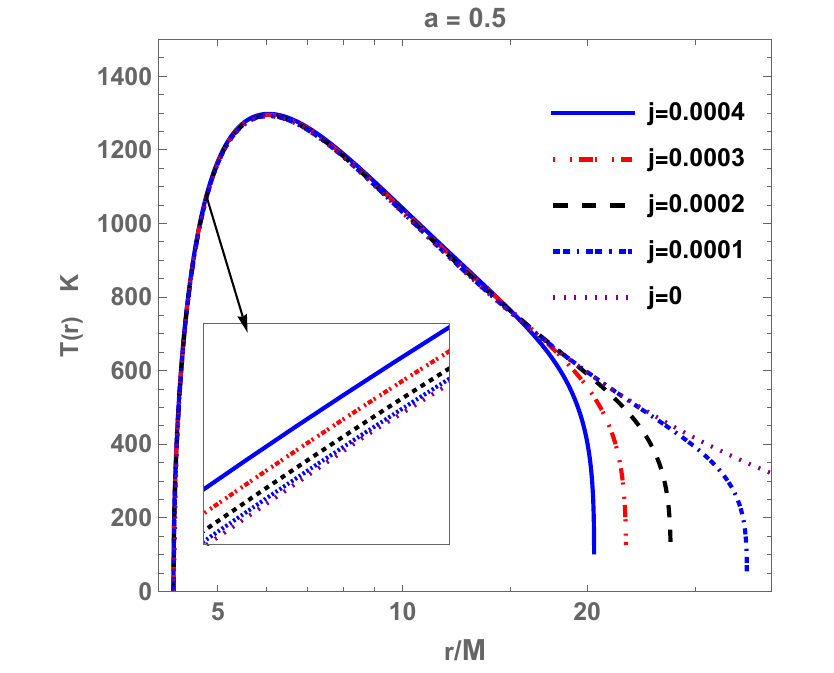}\includegraphics[width=0.34\linewidth]{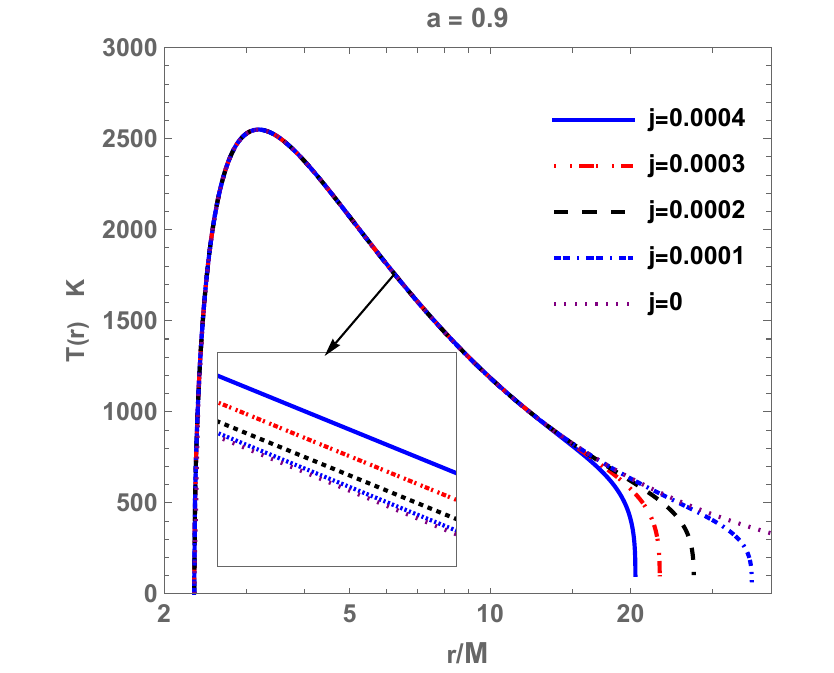} \includegraphics[width=0.32\linewidth]{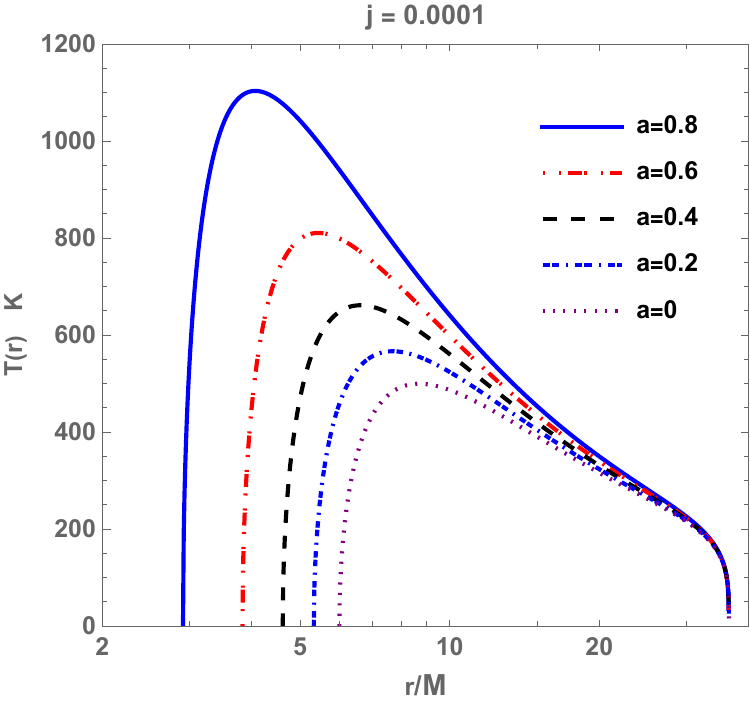}\;\;\;\includegraphics[width=0.32\linewidth]{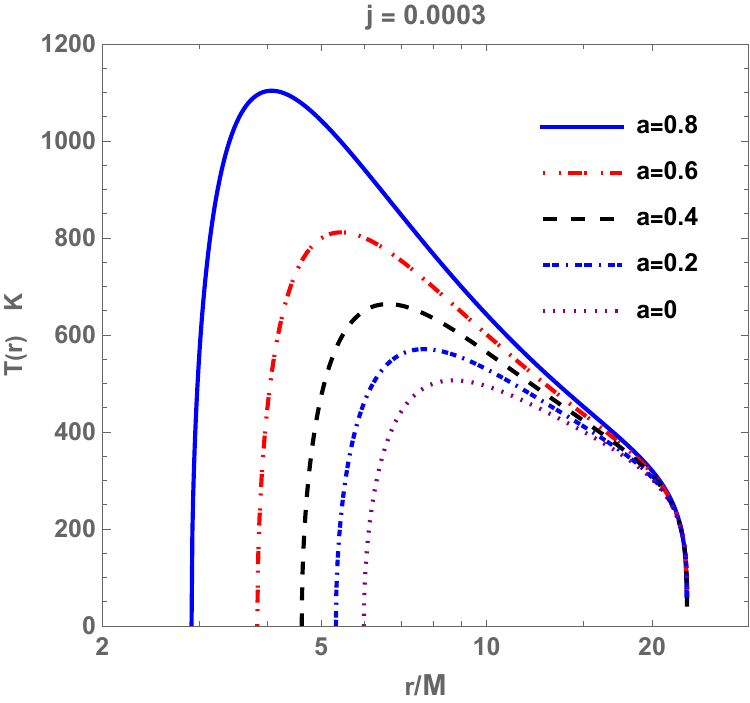}\;\;\;\includegraphics[width=0.32\linewidth]{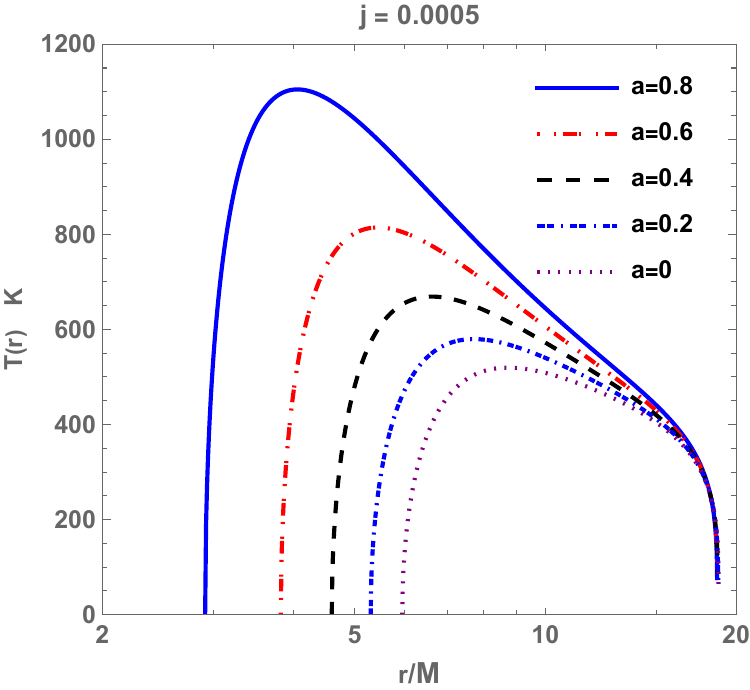}
        \captionsetup{justification=raggedright, singlelinecheck=false}
 	\caption{Variation of the radiation temperature $T(r)$ with the swirling parameter $j$ and the black hole spin parameter $a$  in the thin disk around the rotating swirling black hole. Here, we set $M=1$.}
        \label{T}
\end{figure}
Moreover, as  the black hole spin parameter $a$ increases, the effects of the swirling parameter $j$ on the energy flux decrease.

In the steady-state thin disk model \cite{IBH}, it is generally assumed  that the accreting matter is in thermodynamic equilibrium and the radiation emitted by the disk surface can be considered as perfect black body radiation. In terms of the Stefan-Boltzmann law $T(r)=[K(r)/\sigma]^\frac{1}{4}$, where $\sigma$ is the Stefan-Boltzmann constant, one can analyze the dependence of the radiation temperature on the swirling parameter $j$.
Fig. (\ref{T}) presents the radiation temperature profile as a function of $r$. It is shown that the effects of the swirling parameter $j$ and spin parameter $a$ on the radiation temperature are similar to those on the energy flux $K$. Similarly, we also note that the effects of the swirling parameter are suppressed by the black hole spin parameter $a$.

\begin{figure}[ht]
    \centering
		\includegraphics[width=0.31\linewidth]{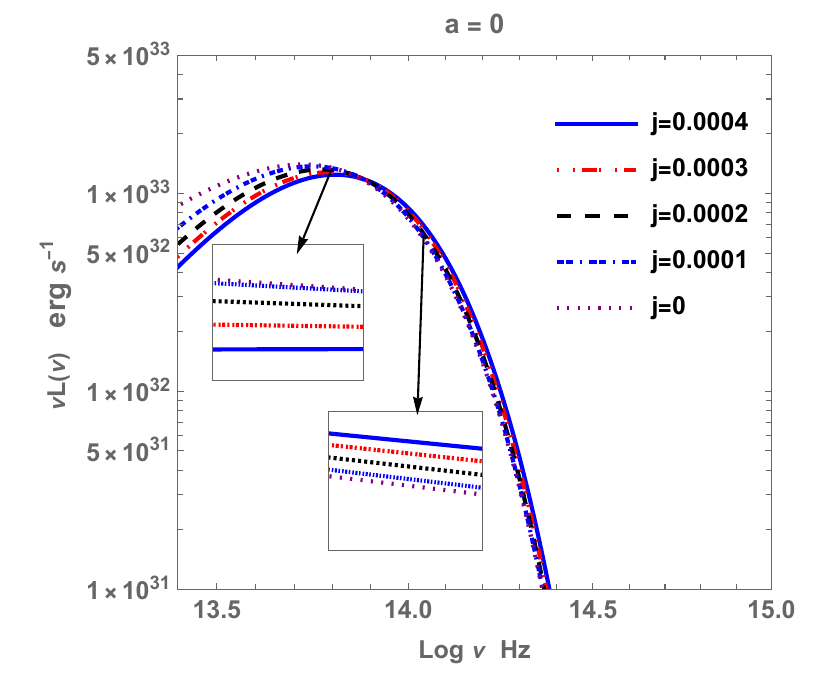} \includegraphics[width=0.31\linewidth]{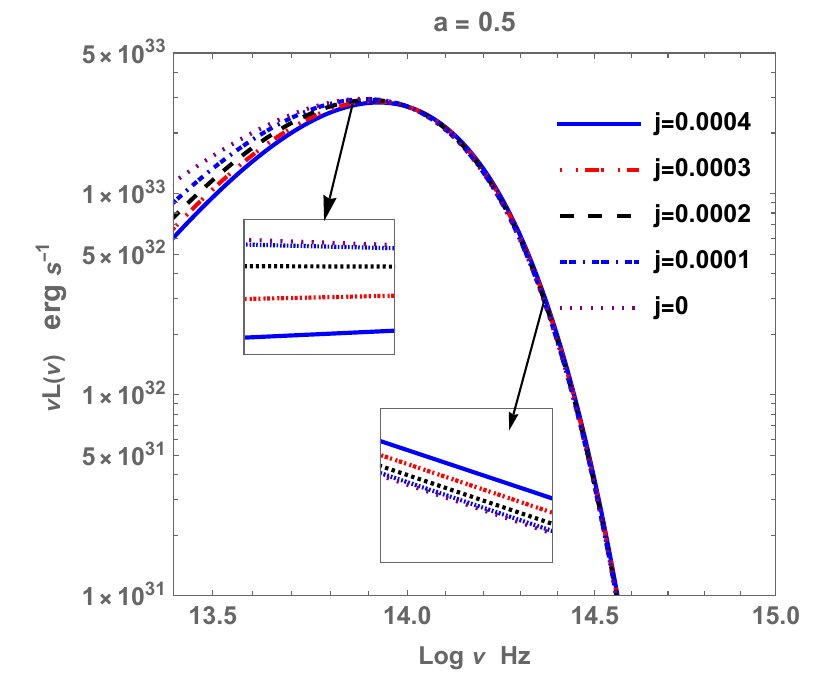} \includegraphics[width=0.31\linewidth]{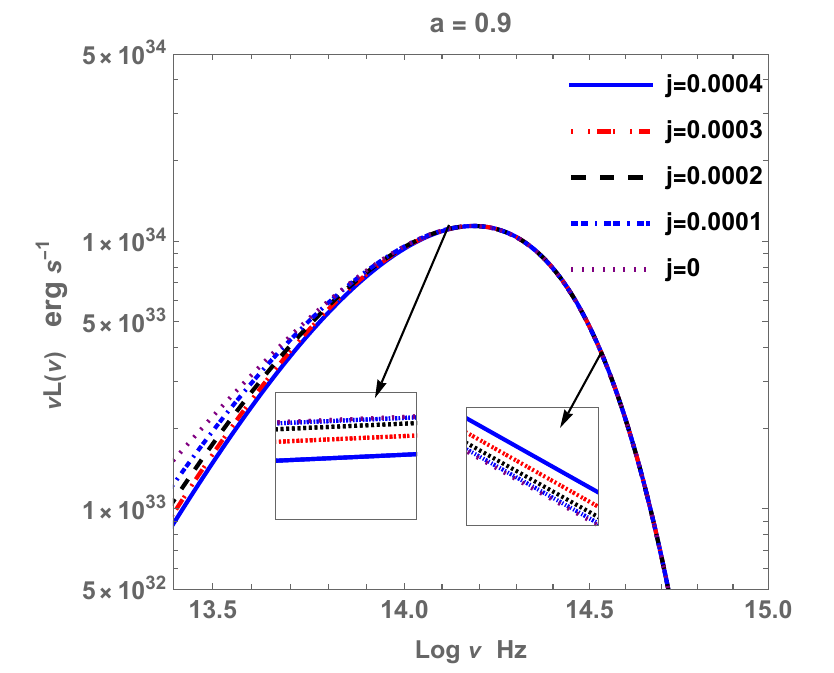}\\
		\includegraphics[width=0.31\linewidth]{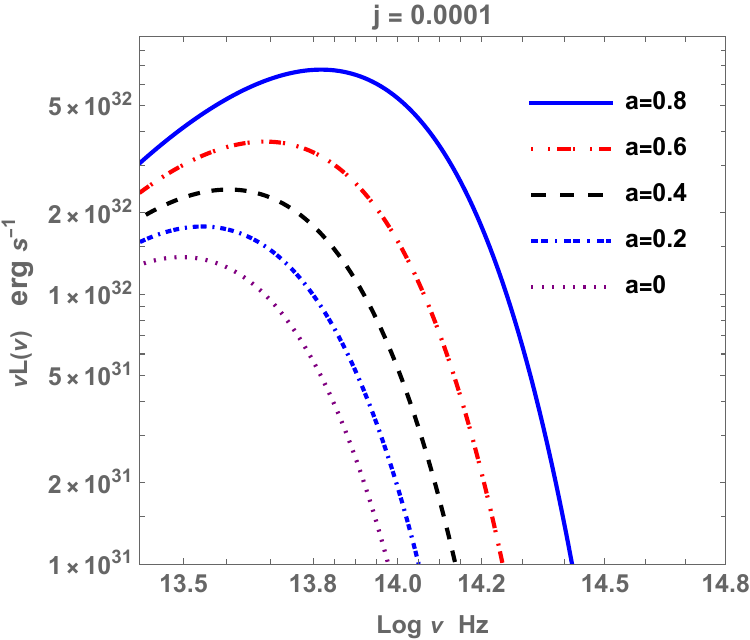} \includegraphics[width=0.31\linewidth]{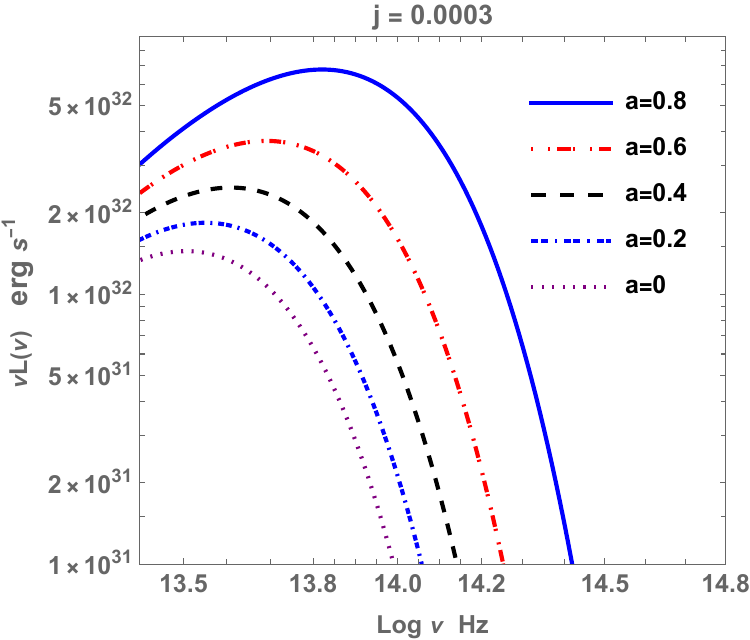}  \includegraphics[width=0.31\linewidth]{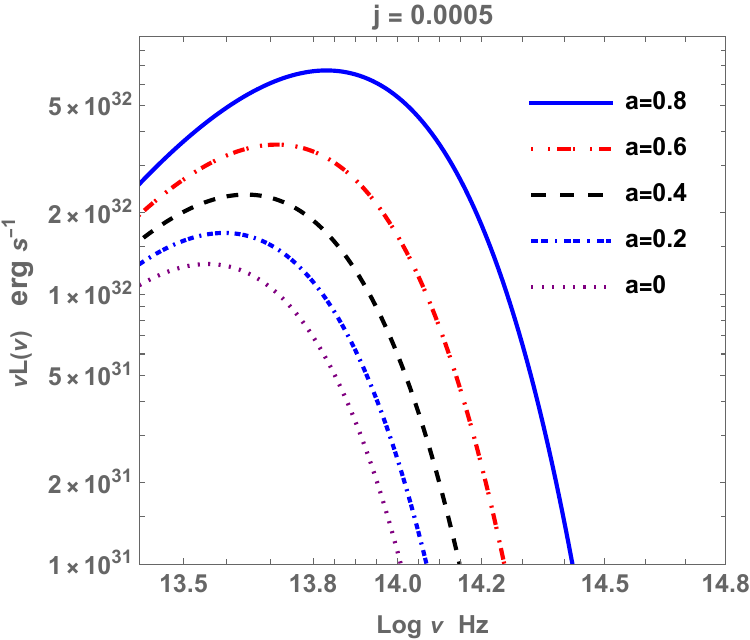}
       \captionsetup{justification=raggedright, singlelinecheck=false}
 	\caption{Variation of the emission spectrum  with the swirling parameter $j$ and the black hole spin parameter $a$  in the thin disk around the rotating swirling black hole. Here, we set $M=1$.}
        \label{vL}
\end{figure}

Considering the thermal black body radiation, the luminosity of the accretion disk can be obtained by \cite{BCO2}
\begin{equation}\label{guangdu}
    \begin{aligned}
        L(\nu)=\frac{16\pi^{2}h\cos\gamma}{c^{2}}\int_{r_{i}}^{r_{out}}\frac{\nu^{3}r}{e^{h\nu/kT(r)}-1} dr.
    \end{aligned}
\end{equation}
Here, $\gamma$ represents the inclination angle of the accretion disk, and $r_{i}$ and $r_{out}$ denote the inner and outer boundaries of the disk, respectively. Based on previous analysis, we restrict our analysis to the case of $\gamma = 0^\circ$ for the sake of convenience. Resorting to numerical methods, we calculate the integral (\ref{guangdu}) and present the spectral energy distribution of the disk radiation in Fig. (\ref{vL}).
It is shown that the larger values of $j$ and $a$ leads to higher cut-off frequencies. However, the presence of $j$ decreases the observed luminosity of the disk at lower frequencies and enhances the observed luminosity only at higher frequencies, which differs from the effect of the black hole spin.

\begin{figure}
 		\includegraphics[width=5cm]{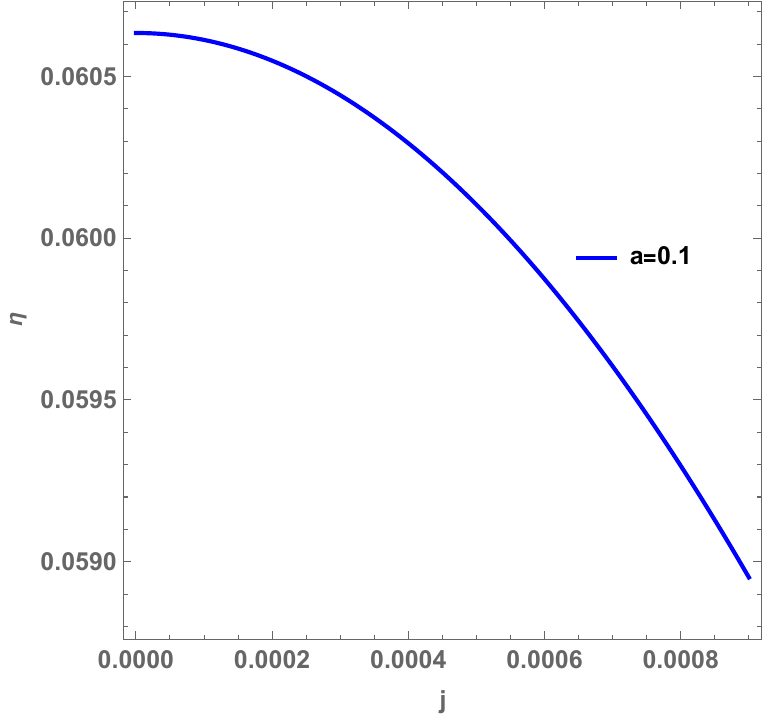}
        \includegraphics[width=5cm]{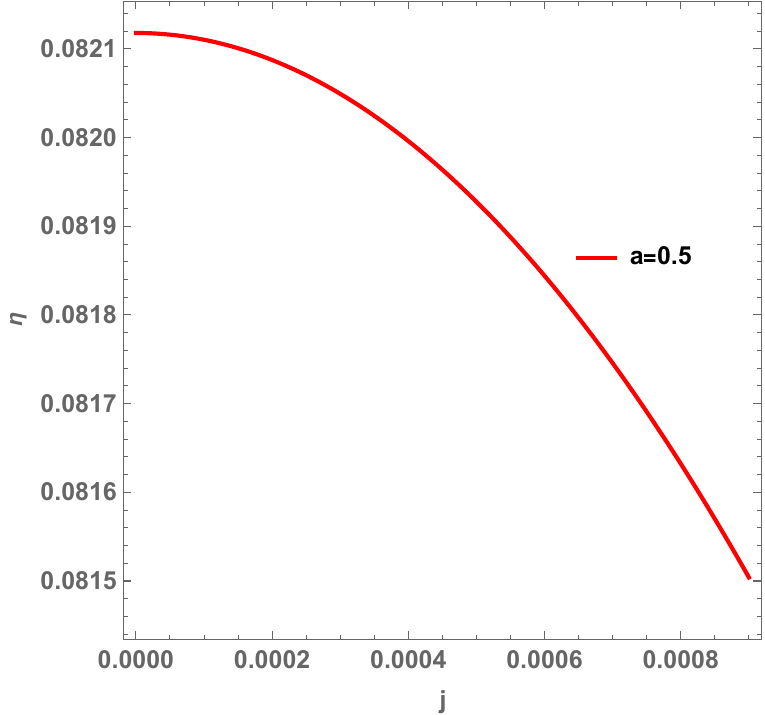}
        \includegraphics[width=5cm]{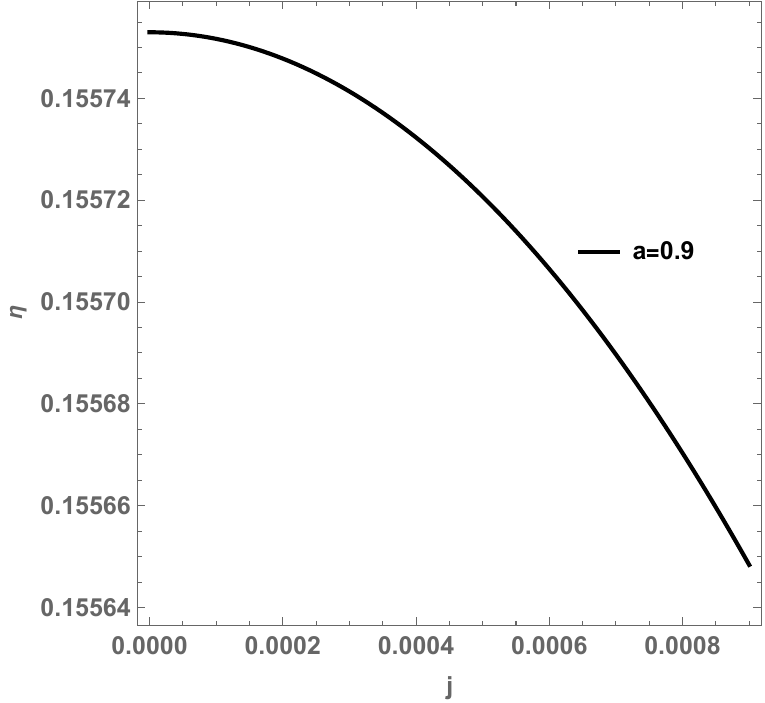}
 \caption{Variation of the efficiency $\epsilon$  with the swirling parameter $j$ and the black hole spin parameter $a$  in the thin disk around the rotating swirling black hole. Here, we set $M=1$.}
 \label{eff}
 \end{figure}

In the accretion process, the efficiency $\epsilon$ is used to  describe the capability of the central object to convert rest mass into outgoing radiation.
Generally, the conversion efficiency can be obtained by the ratio between the rate of the radiant energy of photons escaping from the disk surface to infinity and the mass-energy transfer rate of the central compact object in  mass accretion.
Assuming that all emitted photons can escape to infinity, $\epsilon$ depends solely on the specific energy at the marginally stable orbital radius $r_{ms}$, which is given by
\begin{equation}\label{xiaolv}
    \begin{aligned}
       \epsilon=1-E_{r_{ms}}
    \end{aligned}
\end{equation}
The dependence of the conversion efficiency on the swirling parameters $j$ and the black hole spin parameter $a$ is plotted in Fig. (\ref{eff}), which shows that the conversion efficiency increases with the black hole spin parameter $a$, but decreases with on the swirling parameter $j$.
Moreover, with the increase of the black hole spin parameter $a$, the difference in the conversion efficiency arising from the swirling parameter $j$ decreases.

\section{Summary}

We have studied the properties of the thin accretion disk in the swirling-Kerr black hole background. Our results show that the swirling parameter $j$ imprints on the energy flux, temperature distribution and emission spectra of the disk and  gives rise to some new effects different from those arising from the black hole spin $a$. For a fixed circular orbital radius, the specific energy $E$, specific angular momentum $L$, and angular velocity $\Omega_{\phi}$ of the particle increases with the swirling parameter $j$, but decreases with the black hole spin parameter $a$. In particular, with the increase of the orbital radius,  the angular velocity $\Omega_{\phi}$ in the nonzero $j$ case first decreases and then increases, which is different from that in the pure Kerr black hole case where it decreases monotonically. The marginally stable orbit radius $r_{ms}$ decreases with swirling parameter $j$ both in the prograde case and the retrograde case for different $a$. With the increasing of the swirling parameter $j$, the energy flux $K$ and  radiated temperature $T$ increase in the inner region with the smaller circular orbital radii and decrease in the outer region with the larger circular orbital radii. However, they always increase with the black hole spin $a$. Although the larger values of $j$ and $a$ leads to higher cut-off frequencies, the presence of $j$ decreases the observed luminosity of the disk at lower frequencies and enhances the observed luminosity only at higher frequencies, which differs from the effect of the black hole spin. Moreover, the conversion efficiency increases with the black hole spin parameter $a$ but decreases with the swirling parameter $j$.
We also note that the effects of the swirling parameter are suppressed by the black hole spin parameter $a$, which means that effects of the swirling parameter become more distinct in spacetimes with black holes of lower spin. These results could help us further understand the properties of thin accretion disks and the swirling of the universe background.

\section{\bf Acknowledgments}
This work was  supported by the National Natural Science
Foundation of China under Grant No.12275078, 11875026, 12035005, 2020YFC2201400 and the innovative research group of Hunan Province under Grant No. 2024JJ1006.

\end{document}